\documentclass[onecolumn,modern]{aastex62}
\received{April 24, 2018}
\submitjournal{Astronomy Letters (Pis'ma v Astronomicheskii Zhurnal)}

\shortauthors{Nikolaeva et al.}

\def\x6{M33~X-6}  
\def\xrt{{\it Swift}-XRT} %
\def\nu{{\it NuSTAR}} %
\def\lum{~erg~s$^{-1}$} %
\def\flux{~erg~cm$^{-2}$~s$^{-1}$} %
\usepackage{tabularx}
\begin{document}

\title{Broadband Spectrum of the X-ray Binary \x6\ from \nu\
and \xrt\ Data: An Extragalactic Z-Source?}

\correspondingauthor{S.M. Nikolaeva}
\email{nikolaeva.sm@phystech.edu}

\author{S.M. Nikolaeva}
\affiliation{Space Research Institute, Russian Academy of Sciences, Profsoyuznaya ul. 84/32, Moscow, 117997 Russia}
\affiliation{Moscow Institute of Physics and Technology (State University), Dolgoprudnyi, Moscow obl., 141701 Russia}

\author[0000-0003-2737-5673]{R.A. Krivonos}
\affiliation{Space Research Institute, Russian Academy of Sciences, Profsoyuznaya ul. 84/32, Moscow, 117997 Russia}

\author{S.YU. Sazonov}
\affiliation{Space Research Institute, Russian Academy of Sciences, Profsoyuznaya ul. 84/32, Moscow, 117997 Russia}



\begin{abstract}
We present the results of our study of the X-ray spectrum for the source X-6 in the nearby galaxy M33 obtained for the first time at energies above 10 keV from the data of the \nu\  orbital telescope. The archival \xrt\ data for energy coverage below 3 keV have been used, which has allowed the spectrum of \x6\ to be constructed in the wide energy range 0.3--20 keV. The spectrum of the source is well described by the model of an optically and geometrically thick accretion disk with a maximum temperature of $\sim 2$ keV and an inner radius of $\sim 5\cos^{-1/2}\theta$ km (where $\theta$ is the unknown disk inclination angle with respect to the observer). There is also evidence for the presence of an additional hard component in the spectrum. The X-ray luminosity of \x6\ measured for the first time in the wide energy range 0.3--20 keV is $\sim 2 \times10^{38}$\lum , with the luminosity in the hard 10--20 keV X-ray band being $\sim10$\% of the source’s total luminosity. The results obtained suggest that X-6 may be a $Z$-source, i.e., an X-ray binary with subcritical accretion onto a weakly magnetized neutron star.
\end{abstract}





\section{Introduction} \label{sec:intro}

The first detailed observations of nearby galaxies carried out onboard the Einstein observatory \citep{giacconi79} revealed luminous X-ray sources unrelated to the activity in galactic nuclei characteristic for Seyfert galaxies. Seventeen X-ray sources were detected in the nearby spiral galaxy M33 (Triangle) among which there are a luminous central source with a luminosity of $10^{39}$\lum named X-8 (the ordinal number in the list in right ascension) and several other sources in the disk and spiral arms of the galaxy with a luminosity of $10^{37}$--$10^{38}$\lum\ \citep{long81, markert83, Trinchieri88}. In addition to point sources, an intense extended emission was detected. A ROSAT survey of the galaxy M33 expanded the list of detected sources to 57 \citep{Schulman95,long96} and showed that the extended emission is consistent with the spiral structure of the galaxy closer to its central region. Subsequently, using the archival ROSAT data, \cite{Haberl01} presented a catalog of 184 sources within 50 arcmin of the M33 central region. Succeeding surveys of M33 onboard the modern Chandra and XMM-Newton observatories revealed hundreds of X-ray sources, gradually resolving the extended emission into separate sources in the disk and spiral arms of the galaxy \citep{Pietsch04, Misanovic06, Plucinsky08, Tullmann11}.

\begin{deluxetable*}{cccccc}[t]
\tablecaption{Description of the \nu\ observations\label{meansp1}}
\tablewidth{0pt}
\tablehead{
\colhead{Obs \#} & \colhead{Obs Id} & \colhead{Date} & \colhead{Exposure} & \colhead{FPMA count} &  \colhead{FPMB count} \\
\nocolhead{} & \nocolhead{} & \nocolhead{} & \colhead{ks} & \colhead{rate, 10$^{-2}$ s$^{-1}$} &  \colhead{rate, 10$^{-2}$ s$^{-1}$}}
\startdata
1 & 50310001002 & 2017-03-04 & 108 & 2.54 $\pm$ 0.08 & 2.76 $\pm$ 0.07\\ 
2 & 50310001004 & 2017-07-21 & 99 & 2.96 $\pm$ 0.09 & 3.13 $\pm$ 0.09 \\
\enddata
\end{deluxetable*}

The \nu\ observatory conducted a survey of the galaxy M33 at energies above 10 keV with an angular resolution previously inaccessible for these energies. The observations were performed within the framework of a ``legacy survey'' whose data are open for the astrophysical community\footnote{\url{https://www.nustar.caltech.edu/page/legacy_surveys}}. Thus, it has become possible to carry out a spectral analysis of M33 sources in the hard X-ray band for the first time. For example, \cite{krivonos18} has studied the broadband spectrum of the luminous central source M33 X-8 for the first time, whose luminosity ($\sim 2\times 10^{39}$\lum) allows it to be attributed  to the class of ultraluminous X-ray sources. It turned out that the spectrum could be described as the sum of a standard accretion disk \citep{shakura73} with a temperature of 1~keV and a hard power-law continuum with a high slope ($\Gamma\sim 3$) extending at least to 20 keV. Thus, the ultraluminous X-ray source M33 X-8 turned out to be similar in its spectral properties to the most luminous ($\sim 10^{38}$--$10^{39}$~\lum) and very few X-ray binaries in our Galaxy observed in the so-called very high state. In such objects we may deal with the accretion of matter onto a stellar-mass black hole with a rate that is a significant fraction of the critical one. A further expansion of the sample of well-studied high-luminosity X-ray binaries is needed to test this and alternative hypotheses.

The object of research in this paper, the X--ray source \x6\, is the second brightest one after X--8 in the galaxy M33 \citep{Trinchieri88}, which allowed high-quality spectral data up to 20 KeV to be obtained with the \nu\ telescope. X--ray observations at energies of the standard 2--10~keV X--ray band show that the source is an X--ray binary emitting approximately at the Eddington luminosity for a neutron star (and an appreciable fraction of the Eddington luminosity for a stellar-mass black hole). On the whole, the spectrum of the source is satisfactorily described by the model of an accretion disk with a temperature of 1--2 keV, but there is evidence for an excess emission at energies above 8--~10 keV (\citep{Tullmann11}). Unfortunately, as yet no optical counterpart of the X-ray source \x6\ has been detected, despite the fact that the corresponding field of the galaxy M33 is not very crowded \citep{Tullmann11}.

The distance to the galaxy M33 was measured by several independent methods, but the discrepancy between various measurements reaches $30\%$ \citep[see, e.g.,][]{u2009}. In this paper, to estimate the luminosity of M33 X-6, we used the distance of 817~kpc \citep{freedman2001} previously used to calculate the luminosities of sources from the M33 survey based on Chandra data \citep{Plucinsky08,Tullmann11}.

\section{OBSERVATIONAL DATA} \label{1}
\subsection{NuSTAR} \label{subsec:nustar}
The observations of \x6\ were carried during a survey of the galaxy M33 by the orbital \textit{Nuclear Spectroscopic Telescope Array} \citep[\nu,][]{nustar} at two epochs: in the spring and summer of 2017 (see Table~\ref{meansp1}). \nu\ is the first grazing-incidence telescope operating at energies above 10 keV with the working energy range 3--79 keV. The telescope is equipped with two independent mirror systems, often referred as FPMA and FPMB. The total exposure time for \x6\ was 207 ks.

The lists of photons were preprocessed with the standard \textit{nupipeline} software of the {\sc HEASOFT 6.22}\footnote{\url{https://heasarc.nasa.gov/lheasoft/}} package. As a result, filtered lists of photons were obtained. Subsequently, we extracted the source’s spectra in a circle with a radius of $70''$ and the spectrum of background counts in a ring ${70''<R<125''}$ with the \textit{nuproducts} utility. To satisfy the criterion for applicability of the $\chi^2$ statistic, the source’s spectra were additionally binned with the condition that at least 30 counts were contained in one bin.

\begin{figure*}[ht]
\includegraphics[width=0.5\columnwidth]{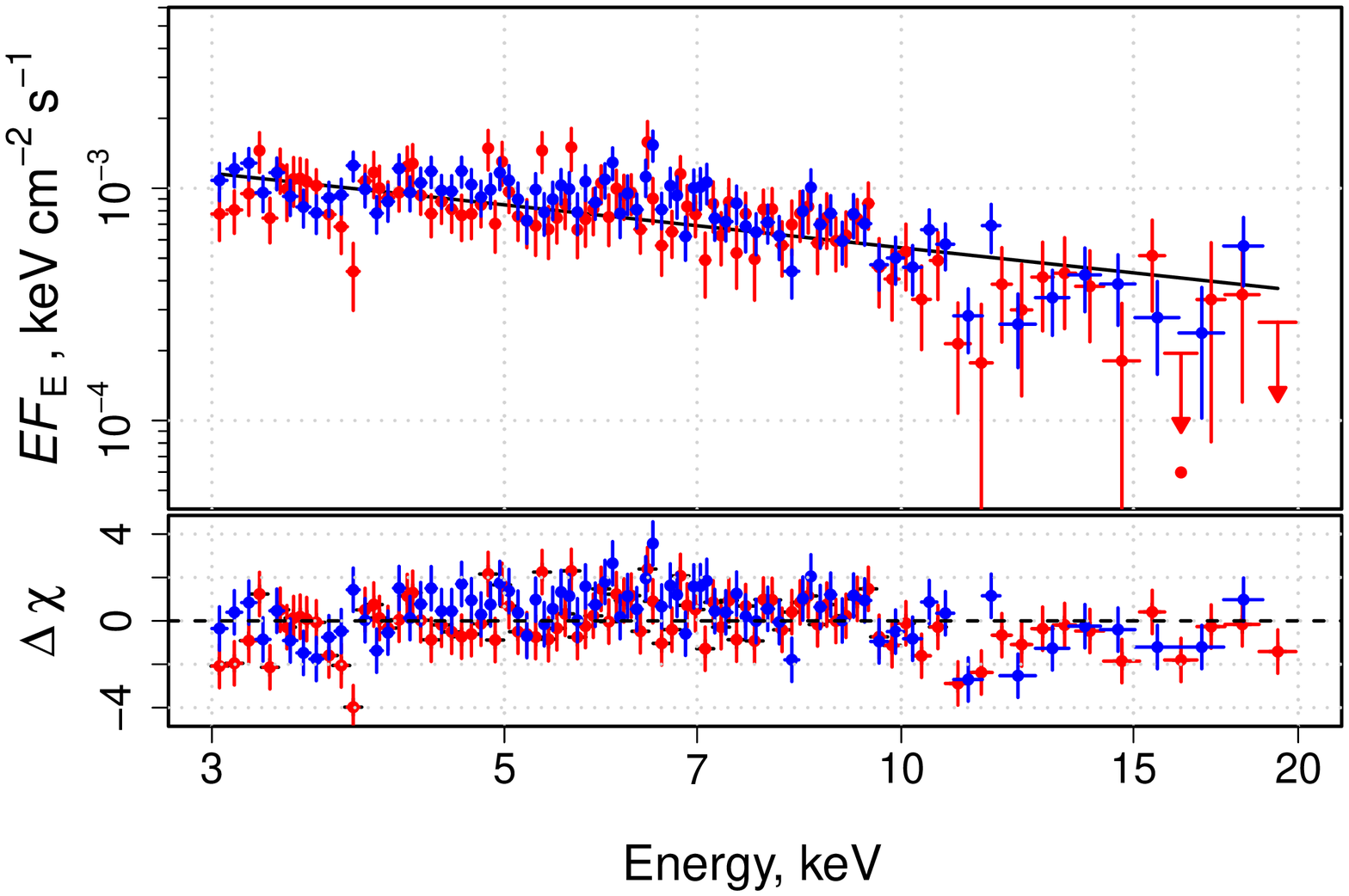}
\includegraphics[width=0.5\columnwidth]{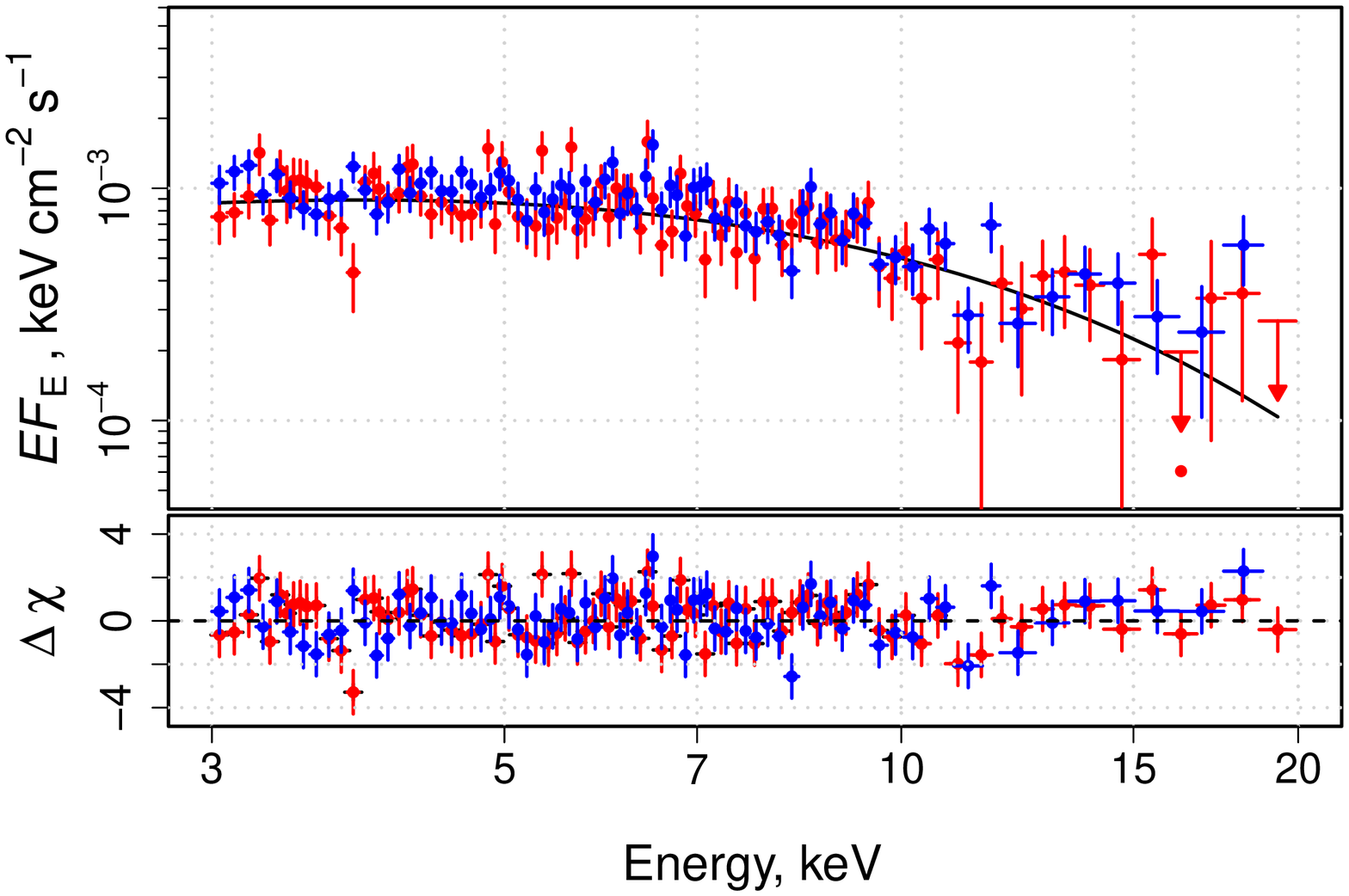}
\caption{\nu\ spectrum of the source in the energy range from 3.0 to 20.0 keV from the \nu\ data for the first epoch of observations. The blue and red dots mark the data from the FMPA and FPMB modules, respectively. The upper limits correspond to 2$\sigma$. The black solid line indicates the fit by a power law ({\texttt phabs$\times$pegpwrl}, left) and a power law with an exponential cutoff ({\texttt phabs$\times$cutoffpl}, right). The statistical deviations of the data from the chosen model are shown on
the bottom panels.} \label{ris:pwlcutpl}
\end{figure*}

A preliminary analysis showed that the source’s spectrum is dominated by the background at energies above 20 keV. For this reason, we did not use the spectral data above this energy in the subsequent analysis. Figure ~\ref{ris:pwlcutpl} shows the spectrum of \x6\ in the energy range 3--20~keV from the spring observations (epoch 1) fitted by a simple power-law model {\texttt phabs$\times$pegpwrl} with the model definition in the energy range 3--10 keV. Note that the \nu\ spectral data are not sensitive enough to the line-of-sight absorption measurement due to the sharp drop in effective area at energies below 5 keV. Therefore, we use the average absorption toward the galaxy M33 $N_H$ = 1.1 $\times$ $10^{21}$ см$^{-2}$ from \cite{kalberla05} for all of the \nu\ spectra. Fitting the model to the data led to an estimate of the power-law slope $\Gamma=2.6\pm0.1$ and the 3--10 keV flux $(1.59 \pm 0.04)\times10^{-12}$\flux. According to the goodness-of-fit criterion $\chi^2_r$/d.o.f. = 1.43/170 (d.o.f. means the number of degrees of freedom), the power-law model describes the data poorly, which manifests itself as large-amplitude variations in the residuals (Fig.~\ref{ris:pwlcutpl}, left).

It can be seen from the results obtained that there is a modest cutoff at high energies in the spectrum. We attempted to describe the spectrum by a power-law model with an exponential cutoff {\texttt phabs$\times$cutoffpl} (Fig.~\ref{ris:pwlcutpl}, right), which improved noticeably the quality of the fit ($\chi^2_r$/d.o.f.=1.08/168). Table~\ref{tab:table2} contains the parameters of the best fit by this model for two epochs of 2017 observations. Note that the cross-normalization constant $C$ means the FPMB/FPMA cross-calibration. The model parameters and the observed flux from the source for the two epochs agree well between themselves. This allows us to conclude that there is no strong variability of \x6\ on 2017 and to combine the data into a single spectrum.

\begin{deluxetable*}{ccCrlc}[b!]
\tablecaption{Best-fit parameters for a power-law model with an exponential cutoff, applied to the \nu\ observations of \x6\ in 2017\label{tab:table2}}
\tablecolumns{6}
\tablewidth{0pt}
\tablehead{
\colhead{Obs Id} &
\colhead{$\Gamma$\tablenotemark{a}} &
\colhead{Cutoff} & \colhead{Flux\tablenotemark{a}, $10^{-12}$} & \colhead{$C$\tablenotemark{b}}& \colhead{$\chi^2_r$, $\chi^2/d.o.f$} \\
\nocolhead{} & \nocolhead{} & \colhead{energy, keV} & \colhead{erg cm$^{-2}$ s$^{-1}$} & \colhead{}
}
\startdata
50310001002 &  1.1$^{+0.05}_{-0.06}$  & 4.32$^{+1.72}_{-0.84}$ & 1.81$^{+0.10}_{-0.07}$ & 1.12 $^{+0.05}_{-0.07}$& 1.08, 181.49/168\\ 
50310001004 & 0.7 $\pm$ 0.4  & 3.92$^{+1.32}_{-0.79}$ & 1.75$^{+0.06}_{-0.08}$ & 1.05$^{+0.06}_{-0.04}$ &  0.83, 157.39/189\\ 
Combined& 0.9 $\pm$ 0.3 & 4.07$^{+0.86}_{-0.56}$ & 1.82$^{+0.05}_{-0.07}$ & 0.95$^{+0.04}_{-0.03}$ & 0.97, 351.88/361 \\ \enddata
\tablenotetext{a}{The absorbed (observed) 3--20~keV flux from the source.}
\tablenotetext{b}{For individual observations the coefficient C means the relative normalization between the FPMA and FPMB observations.}
\end{deluxetable*}

\subsection{Swift-XRT} \label{xrt}
To extend the spectral data to energies below 3~keV, we analyzed the archival data of the Neil Gehrels observatory \citep[\xrt,][]{gehrels04}. We selected the XRT observations in which \x6\ was displaced from the optical axis by no more than $5'$ (see Table~\ref{meansp}), which rules out a strong distortion of the point spread function \cite[see][]{moretti05}. The derived data sample effectively includes two epochs of observations, 2007--2009 and 2015--2016, with a total exposure time of 38~ks, which allows a high quality spectrum to be obtained for a sufficiently bright source (which \x6\ is). Using the \xrt\ data processing software provided by the UK Swift Science Data Centre\footnote{\url{http://www.swift.ac.uk}}, we obtained data for our spectral analysis. In order for the counting statistics to be applicable when analyzing the data (\texttt{cstat} в {\sc xspec}), the \xrt\ spectra were additionally binned by 3 and 7 counts in each bin for the data collection 1--10 and 11, respectively. By analogy with the analysis of the \nu\ data, we fitted the individual \xrt\ spectra by the simple power-law model with line-of-sight absorption \texttt{phabs$\times$powerlaw}. The parameters of this model are given in Table~\ref{meansp3}.

\begin{deluxetable*}{cccccc}[b!]
\tablecaption{Description of the \xrt\ observations\label{meansp}}
\tablecolumns{6}
\tablewidth{0pt}
\tablehead{
\colhead{Obs \#} &
\colhead{Obs ID} &
\colhead{Date} & 
\colhead{Exposure, s} & 
\colhead{Angular offset}& 
\colhead{Count rate} \\
\nocolhead{} & \nocolhead{} & \nocolhead{} & \nocolhead{} & \colhead{from optical axis, $'$} & \colhead{s$^{-1}$}
}
\startdata
 1 & 00031041001 & 2007-12-26 & 2923 & 2.35 & 4.3 $\pm$ 0.4   \\ 
 2 & 00031327001 & 2009-01-20  & 4737 & 2.63  & 5.8 $\pm$ 0.4  \\ 
 3 & 00031327002 & 2009-01-28 & 3646 & 0.48  &  3.6 $\pm$ 0.3 \\
 4 & 00031327003 & 2009-02-10 & 4513 & 2.14  &4.9 $\pm$ 0.3  \\ 
 5 & 00031327004 & 2009-02-24 & 5250 & 2.01  & 6.2 $\pm$ 0.3 \\ 
 6 & 00034202001 & 2015-12-01 & 4737 & 2.63    & 5.4 $\pm$ 0.7  \\ 
 7 & 00034205001 & 2015-12-03 & 2990 & 2.61   &   5.6 $\pm$ 0.4 \\  
 8 & 00034205002 & 2015-12-07 & 2693 & 1.62 & 5.9 $\pm$ 0.5    \\
 9 & 00034205003 & 2015-12-09 & 2981 & 1.98   &   5.3 $\pm$ 0.4  \\  
 10 & 00034205004 & 2016-06-10  & 2933 & 4.36   &   5.9 $\pm$ 0.5 \\  
11 & 00034205005 & 2016-06-15 & 990 & 3.79  & 5.4 $\pm$ 0.7    \\ 
\enddata
\end{deluxetable*}

\begin{figure*}
\begin{center}
\includegraphics[scale=0.55]{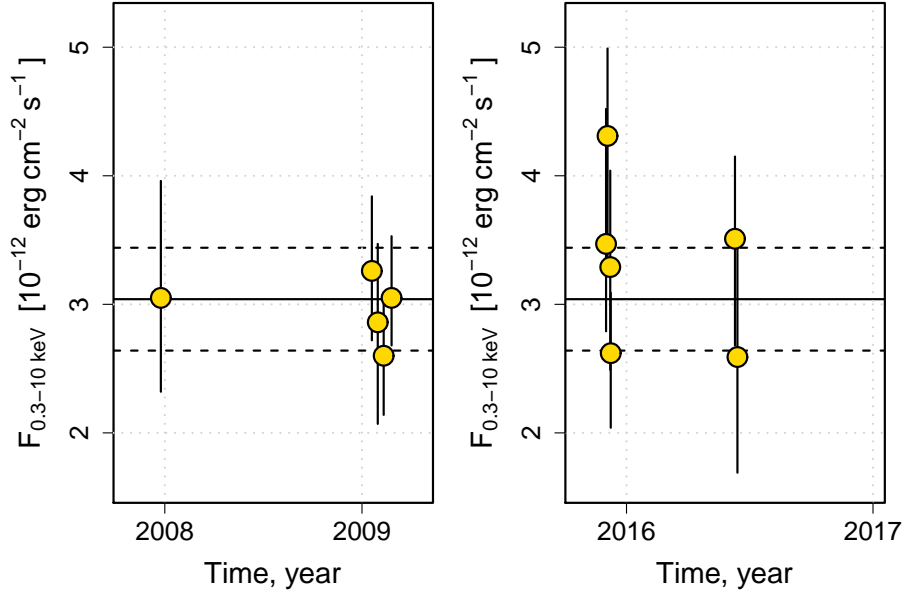}
\caption{Light curve for two epochs of \xrt\ observations in the energy range 0.3--10~keV. The yellow circles represent the observed (absorbed) flux from the source for each individual observation. The horizontal solid line indicates the observed flux for the combined spectrum ($3.04\pm0.33)\times 10^{-12}$\flux\ with its measurement error (dashed lines).}
\label{ris:lc}
\end{center}
\end{figure*}

The parameters of the best fit to the spectra of \x6\ by the power-law model agree well between themselves. Figure~\ref{ris:lc} shows the source’s X-ray light curve from which it can be seen that there was no noticeable variability in the period of observations 2007--2016. This, as with the \nu\ data, allows us to combine the \xrt\ data into a single spectrum. This spectrum constructed in the energy range 0.3-–8 keV is well fitted ($\chi^2_r$/d.o.f.=0.98/277) by a power-law model with a spectral index $\Gamma=1.5\pm0.1$, observed flux $(2.7\pm0.2)\times10^{-12}$\flux\ (0.3--10 keV), and line-of-sight absorption $N_{\rm H} = (1.1 \pm 0.4) \times10^{21}$~cm$^{-2}$. The absorption column density estimate agrees with the column density of interstellar matter in our Galaxy in this direction \citep{kalberla05}.

\begin{deluxetable*}{cccccc}[b!]
\tablecaption{Parameters of the best fit to the \xrt\ data by a power law with an absorption column\label{meansp3}}
\tablecolumns{5}
\tablewidth{0pt}
\tablehead{
\colhead{Obs \#} &
\colhead{$\Gamma$} &
\colhead{$N_{\rm H}$} & 
\colhead{Flux\tablenotemark{a}, $10^{-12}$} & 
\colhead{$\chi^2_{\rm r}$/d.o.f.}  \\
\colhead{} & \nocolhead{} & \colhead{$10^{22}$ cm$^{-2}$} & \colhead{erg cm$^{-2}$ s$^{-1}$} & \colhead{} 
}
\startdata
1 & 1.38$^{+0.43}_{-0.30}$ &  0.09$^{+0.14}_{-0.07}$ & 3.05$^{+0.91}_{-0.73} $ & 1.12/14\\ 
2 & 1.62$^{+0.34}_{-0.29}$ & 0.15$^{+0.09}_{-0.08}$ &3.26$^{+0.58}_{-0.54}$& 1.28/34\\  
3 & 1.11$^{+0.48}_{-0.20}$ & 0.03$^{+0.17}_{-0.01}$ &2.86$^{+0.61}_{-0.79} $& 0.77/15 \\  
4 & 1.77$^{+0.35}_{-0.28}$ & 0.19$^{+0.11}_{-0.11}$& 2.60$^{+0.46}_{-0.46}$ & 0.93/27 \\  
5 & 1.75$^{+0.29}_{-0.22}$ & 0.16$^{+0.10}_{-0.08}$& 3.05$^{+0.48}_{-0.37}$  & 1.24/41\\  
6 & 1.45$^{+0.40}_{-0.32}$ & 0.12$^{+0.14}_{-0.09}$&3.47$^{+1.05}_{-0.68}  $ &0.84/22\\  
7 & 0.90$^{+0.39}_{-0.14}$ & 0.01$^{+0.11}_{-0.01}$ & 4.31$^{+0.68}_{-1.09} $& 0.93/20\\  
8 & 1.52$^{+0.43}_{-0.28}$ & 0.09$^{+0.11}_{-0.07}$& 3.29$^{+0.75}_{-0.80}  $  & 1.00/19\\  
9 & 1.70$^{+0.53}_{-0.31}$& 0.15$^{+0.14}_{-0.10}$ & 2.62$^{+0.47}_{-0.58} $ &  0.92/19\\  
10 & 1.30$^{+0.45}_{-0.18}$ & 0.03$^{-0.02}_{+0.12}$ & 3.51$^{+0.64}_{-0.83}$  & 1.20/21\\  
11 & 1.79$^{+0.75}_{-0.47}$ & 0.13$^{-0.09}_{+0.25}$ &2.59$^{+0.92}_{-0.90}$   &0.86/17\\ 
\enddata
\tablenotetext{a}{The absorbed 0.3--10~keV flux.}
\end{deluxetable*}

\subsection{Broadband Spectrum} \label{sec:broad-band}
\begin{figure*}
\begin{center}
\includegraphics[scale=0.55]{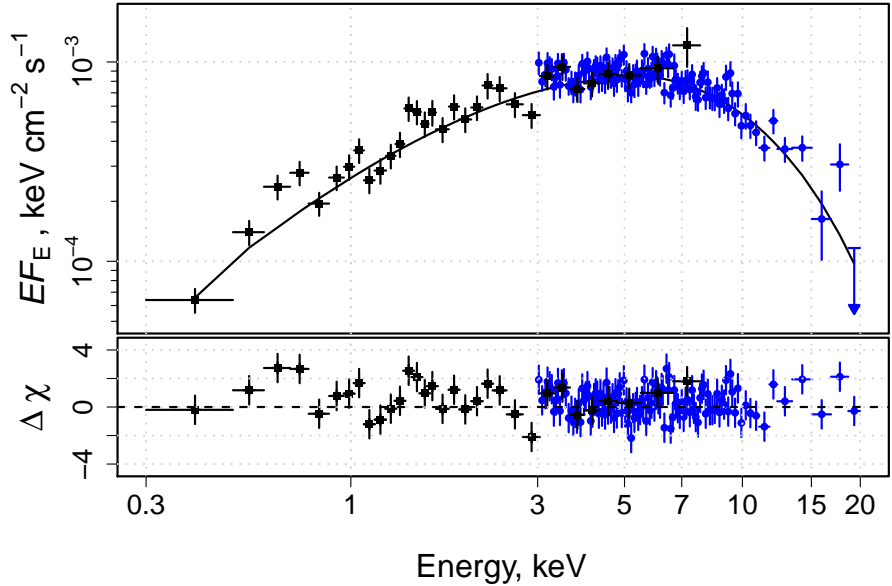}
\caption{Broadband spectrum of \x6\ from the \nu\ and \xrt\ data indicated by the blue circles and black squares, respectively. The fit by the model \texttt{const$\times$phabs$\times$cutoffpl} is indicated by the black solid line. The statistical deviations of the data from the chosen model are shown on the panel below. The upper limits are represented by 2$\sigma$ errors.}
\label{ris:all-cutoffpl}
\end{center}
\end{figure*}

Figure~\ref{ris:all-cutoffpl} shows \x6\ broadband spectrum from 0.3 to 20 keV that was obtained by combining the \nu\ (3--20~keV) and \xrt\ (0.3--10~keV) data. The spectrum was fitted by a powerlaw model with an exponential cutoff and absorption. Table~\ref{tot_app} presents the parameters of the optimal model.

\begin{figure*}
\begin{center}
\includegraphics[scale=0.70]{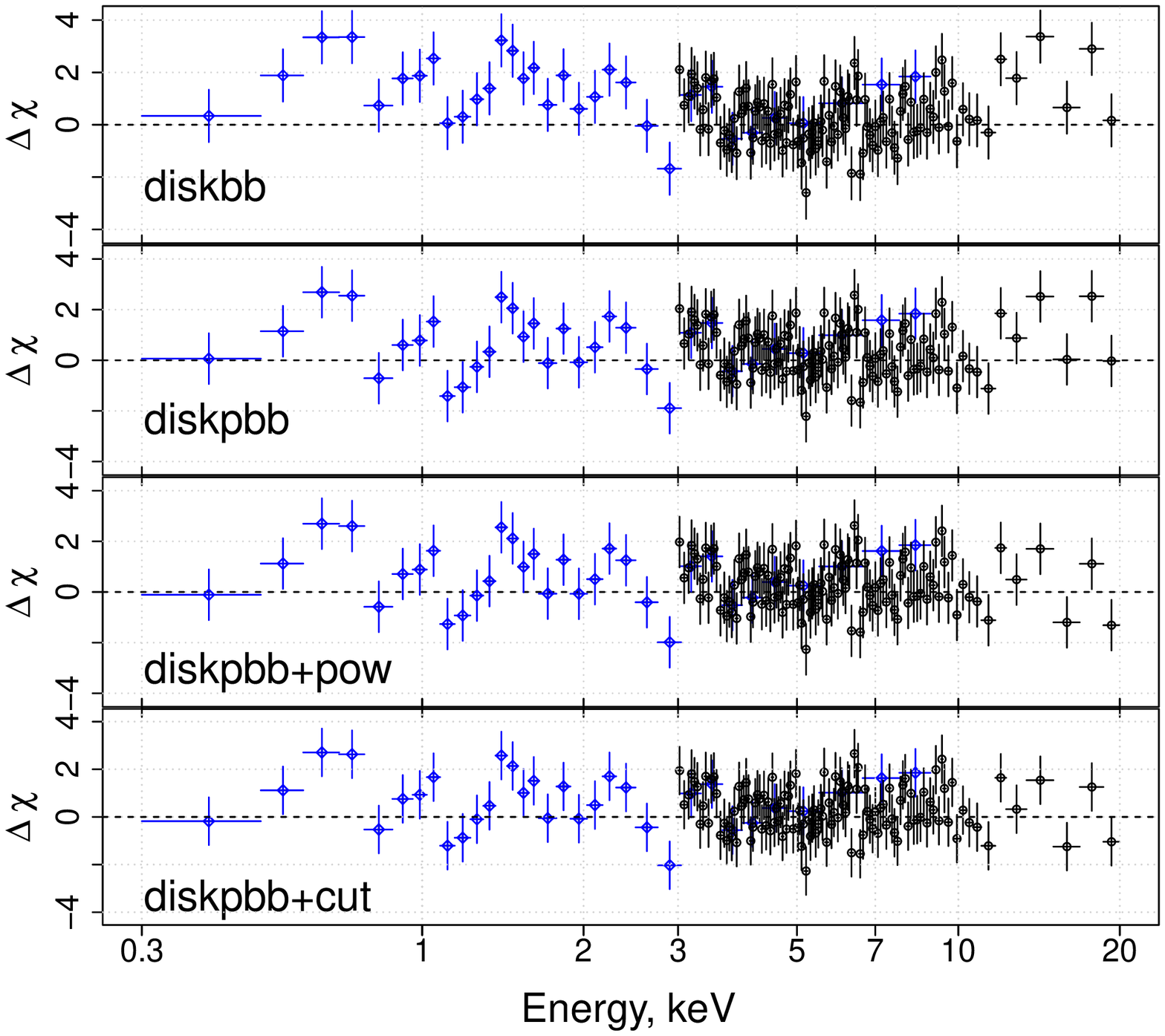}
\caption{Statistical deviations of the broadband spectrum from the chosen models (from top to bottom):
\texttt{phabs$\times$diskbb, phabs$\times$diskpbb }\c, \texttt{phabs$\times$(diskpbb+power-law)} and \texttt{phabs$\times$(diskpbb+cutoffpl)}, where the
blue and black circles mark the \nu\ and \xrt\ data, respectively.}
\label{ris:models_resid}
\end{center}
\end{figure*}

We also estimated the absorbed and unabsorbed fluxes from the source in various energy ranges (Table~\ref{tot_app}); the absorbed flux agrees with the unabsorbed one, within the measurement error limits, due to weak absorption at energies 3.0--20~keV. Assuming the distance to the object to be 817 kpc, we can estimate its luminosity at various energies: $L(3-20 \text{ keV}) = (1.39\pm 0.03)\times 10^{38}$\lum\ and $L(0.3-10 \text{keV}) = (2.05\pm 0.04)\times 10^{38}$\lum. Consequently, the source emits at the Eddington limit in the case of a neutron star and an appreciable fraction of the
Eddington limit in the case of a black hole as a compact object.

As has been shown above in Sections \ref{subsec:nustar} and \ref{xrt}, the source exhibited no strong variability over several years according to the \xrt\ data and during 2017 according to the \nu\ observations. The relative normalization coefficient between the \xrt\ and \nu\ data is $0.86\pm 0.07$, which agrees with the expected discrepancy of 10\% (\cite{madsen15}). Since the source was not highly variable, below we fix the relative normalization between \xrt\ and \nu\ at unity.

Then, we attempted to fit the broadband spectrum of \x6\ by other simple, but more physically motivated models that are commonly used in describing the spectra of X-ray binaries: \texttt{phabs$\times$diskbb} -- the standard multi-temperature accretion disk blackbody model \citep{shakura73}; \texttt{phabs$\times$diskpbb} -- the modified accretion disk model \citep{mineshige94}, in which the slope of the power-law temperature distribution $\propto r^{-p}$ over the disk is a free parameter ($p=0.75$ in the case of a standard disk); \texttt{phabs$\times$(diskpbb+pow)} -- the disk model with an additional power-law component; and \texttt{phabs$\times$(diskpbb+cutoffpl)} -- a similar model with a possible power-law continuum cutoff at high energies. Table~\ref{meansp2} gives the parameters of the optimal models, while Fig.~\ref{ris:models_resid} shows the statistical residuals for each model, respectively.

\begin{table*}[h!]
\centering
\caption{Parameters of fitting of broadband spectrum (0.3--20~keV) by power law model with high energy exponential cutoff.}\label{tot_app} 
\vspace{5mm} 
\begin{tabular}{lcc} \hline \hline
Parameter & Units & Value \\ \hline
$N_H$ &$10^{21}$ cm$^{-2}$ &0.20$^{+0.37}_{-0.17}$    \\  
$\Gamma$& &0.6$^{+0.2}_{-0.1}$ \\ 
$E_{\rm cutoff}$ & keV &3.5$^{+0.4}_{-0.3}$ \\  
$C^a$ & & $0.86\pm0.07$\\
\hline
\multicolumn{3}{c}{Flux, 0.3--10~keV} \\
\multicolumn{1}{r}{(absorbed)} & $10^{-12}$ erg cm$^{-2}$ s$^{-1}$ &2.53$\pm0.05$\\  
\multicolumn{1}{r}{(unabsorbed)} &$10^{-12}$ erg cm$^{-2}$ s$^{-1}$  &2.57$\pm0.05$\\ 
\hline
\multicolumn{3}{c}{Flux, 3--20~keV} \\
\multicolumn{1}{r}{(absorbed)} &$10^{-12}$ erg cm$^{-2}$ s$^{-1}$  &1.74$\pm0.04$  \\  
\multicolumn{1}{r}{(unabsorbed)} &$10^{-12}$ erg cm$^{-2}$ s$^{-1}$  &1.74$\pm0.04$  \\  
\hline
\multicolumn{3}{c}{Flux, 0.3--20~keV} \\
\multicolumn{1}{r}{(absorbed)} &$10^{-12}$ erg cm$^{-2}$ s$^{-1}$  &$2.83\pm0.05$\\ 
\multicolumn{1}{r}{(unabsorbed)} &$10^{-12}$ erg cm$^{-2}$ s$^{-1}$  &$2.88\pm0.05$\\ 
\hline
 $\chi^2_r$, $\chi^2/d.o.f$ & & 0.99, 635.13/641\\ \hline
\end{tabular} 
\begin{flushleft}
$^{a}C$ is a cross-correlation constant between \xrt\ и \nu\ observations.
\end{flushleft}
\end{table*}

The first model describes poorly the data ($\chi^2_r/d.o.f=1.05/642$), leaving a strong excess of photons at energies above 10 keV. The second model improves significantly the fit ($\chi^2_r/d.o.f=1.00/641$) with a temperature profile $p=0.64^{+0.03}_{-0.01}$, but a noticeable data excess at high energies is still present. Adding a power-law component to the disk model does not change dramatically the statistic $\chi^2_r/d.o.f.= 0.99/639$ ($\Delta\chi^2_r=8$ when reducing the number of degrees of freedom by 2), but on the third panel in Fig.~\ref{ris:models_resid} it can be seen how the excess of photons in the right part of the spectrum became smaller. The model consisting of disk emission and a power-law continuum with an exponential cutoff describes satisfactorily the data, restricting the cutoff at energy $\sim$8 keV, but it does not improve the $\chi^2$ statistic compared to the model without an exponential cutoff.

\begin{table*}[h!]
\centering
\caption{Parameters of the best fit to the broadband spectrum
(0.3--20~keV) by simple models.}\label{meansp2} 
\vspace{5mm} 
\begin{tabular}{c c c} \hline \hline
Parameter & Units & Value \\ \hline
\multicolumn{3}{c}{Model 1: phabs$\times$diskbb}  \\ \hline 
 $N_{\rm H}$&$10^{21}$ cm$^{-2}$& $<0.1$  \\
 $kT$ & keV& 2.18$^{+0.02}_{-0.06}$\\
 $N_{\rm diskbb}$ &10$^{-3}$&5.6$^{+0.8}_{-0.2}$ \\
 $R\cos^{1/2}\theta$ & km & 6.1$^{+0.4}_{-0.1}$\\
 $\chi^2_r$, $\chi^2/d.o.f$ & & 1.05, 672.28/642\\ \hline
\multicolumn{3}{c}{Model 2: phabs$\times$diskpbb}  \\ \hline 
 $N_H$&$10^{21}$ cm$^{-2}$& 0.6$\pm 0.3$ \\
 $kT$ & keV& 2.43$^{+0.09}_{-0.08}$\\
 $p$ & & 0.64$^{+0.03}_{-0.01}$\\
 $N_{\rm diskpbb}$ &10$^{-3}$&2.4$^{+0.6}_{-0.4}$ \\
 $R\cos^{1/2}\theta$ & km & 4.0$^{+0.5}_{-0.3}$ \\
 $\chi^2_r$, $\chi^2/d.o.f.$ & &1.00, 640.17/641\\ \hline
\multicolumn{3}{c}{ Model 3: phabs$\times$(diskpbb+pow)}  \\ \hline 
 $N_H$&$10^{21}$ sm$^{-2}$& 0.4$^{+0.5}_{-0.2}$ \\
 $kT$ & keV& 2.23$^{+0.14}_{-0.13}$ \\
 $p$ & & 0.67$^{+0.02}_{-0.03}$\\
 $N_{\rm diskpbb}$ &10$^{-3}$& 3.5$^{+1.3}_{-1.0}$ \\
  $R\cos^{1/2}\theta$ & km & 4.8$^{+0.9}_{-0.7}$\\
 $\Gamma$ & & -0.12$^{+3.9}_{-0.49}$\\
 $N_{\rm pow}$ &10$^{-7}$&2.9$^{+28.7}_{-2.2}$ \\
  $\chi^2_r$, $\chi^2/d.o.f$ & & 0.99, 632.49/639 \\ \hline
\multicolumn{3}{c}{ Model 4: phabs$\times$(diskpbb+cutoffpl)}  \\ \hline 
 $N_H$&$10^{21}$ см$^{-2}$& 0.39$^{+0.06}_{-0.03}$ \\
  $kT$ & keV& 2.16$^{+0.08}_{-0.11}$ \\
 $p$ & & 0.68$^{+0.04}_{-0.05}$\\
 $N_{\rm diskpbb}$ &10$^{-3}$& 4.3$^{+1.2}_{-0.4}$ \\
 $R\cos^{1/2}\theta$ & km & 5.4$^{+0.7}_{-0.2}$\\
 $\Gamma$ & & -0.58$^{+0.06}_{-0.04}$\\
 Cutoff energy &keV& 8.1$^{+1.4}_{-1.5}$\\
 $N_{\rm cutoffpl}$ &10$^{-7}$&7.78$^{+0.60}_{-0.77}$ \\
 $\chi^2_r$, $\chi^2/d.o.f$ & & 0.99, 632.16/638 \\
\hline
\end{tabular} 
\end{table*}

The inner radius of the accretion disk determined by modeling the spectrum turns out to be $\sim 5\cos^{-1/2}\theta$~km. Such a small radius suggests that the relativistic compact component in this binary can be a neutron star rather than a black hole, although there is a large uncertainty in the disk inclination angle $\theta$ with respect to the observer. In the case of a neutron star the measured X-ray luminosity of \x6\, $\sim 2\times 10^{38}$\lum, implies that accretion occurs approximately with the critical rate.

\begin{figure*}
\begin{center}
\includegraphics[scale=0.70]{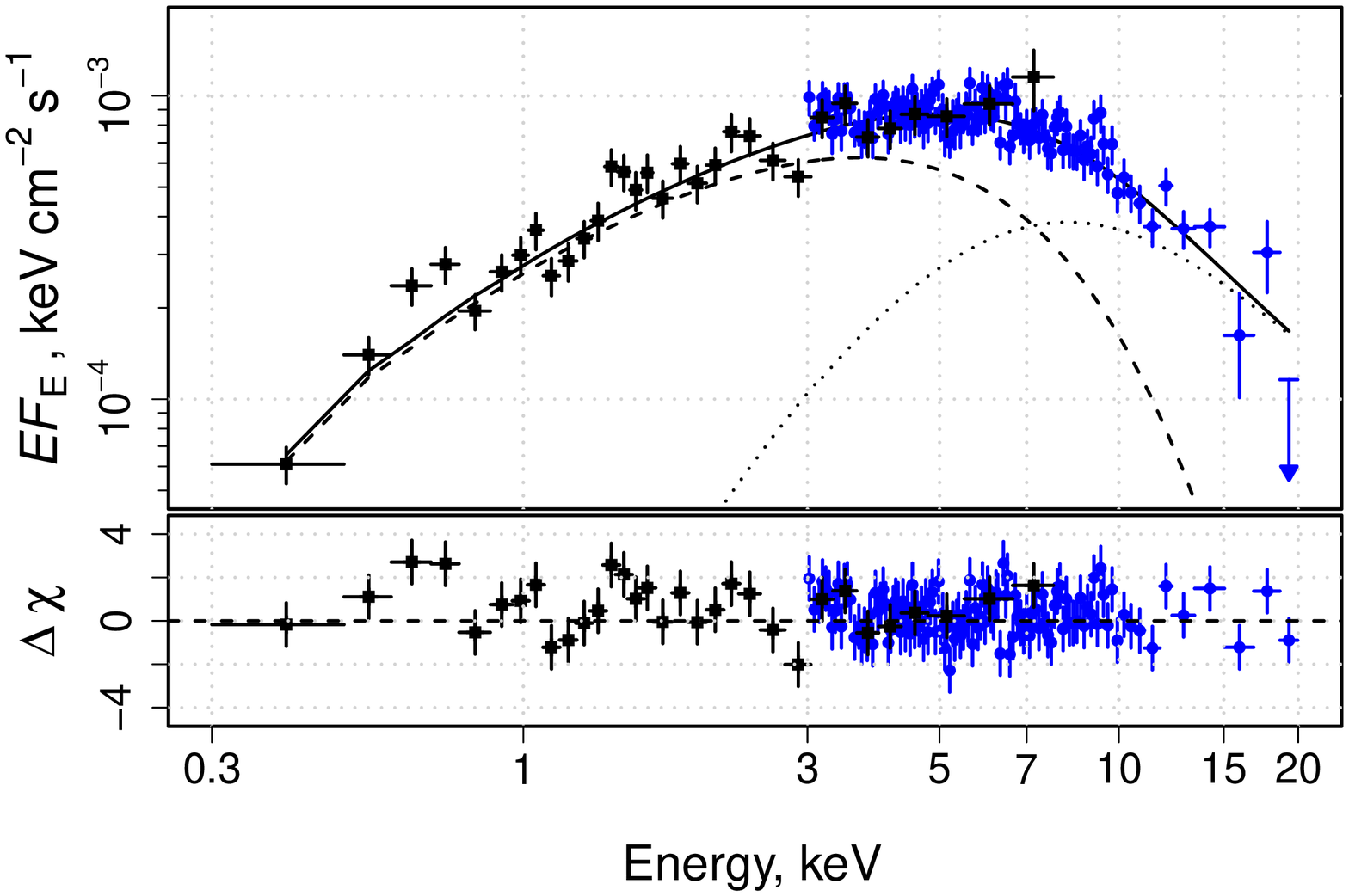}
\caption{Broadband spectrum of \x6\ from the \nu\ and \xrt\ data indicated by the blue circles
and black squares, respectively. The fit by the model \texttt{diskpbb+CompTT} is indicated by the black solid line. The emission components associated with the accretion disk and Comptonization in the hot corona are indicated by the dashed and dotted lines, respectively. The upper limits correspond to 2$\sigma$. The data residuals relative to the model are shown on the bottom panel.}
\label{ris:t_all_data}
\end{center}
\end{figure*}

Next, we considered the model in which the component \citep[\texttt{compTT,}][]{titarchuk94} associated with Comptonization in a cloud of hot electrons was added to the thermal accretion disk emission (\texttt{diskpbb}); for  simplicity, the temperature of the incident photons for Comptonization was taken to be equal to the temperature of the disk at its inner boundary. This model describes the data as well ($\chi^2_r/d.o.f.$=0.99/638, Fig.~\ref{ris:t_all_data}) and models 4 and 5 considered above; the optical depth of the hot corona turns out to be approximately equal to unity, while the electron temperature of the corona is $kT_{\rm e} = 12\pm3$~keV (Table~\ref{last}).

\section{Discussion and conclusions}
In this paper we have constructed a broadband (0.3--20~keV) spectrum of the X-ray source \x6\ in the nearby galaxy M33 for the first time from the \xrt\ (0.3--10~keV) and \nu\ (3--20~keV) data. 

\begin{table*}[h!]
\centering
\caption{Parameters of the best fit to the broadband spectrum
by a combination of accretion disk emission \texttt{diskpbb}
and Comptonized hot corona emission \texttt{compTT}} 
\label{last} 
\vspace{5mm} 
\begin{tabular}{c c c} \hline \hline
Parameter & Units & Value \\ \hline
\multicolumn{3}{c}{Model 5: \texttt{phabs$\times$(diskpbb+compTT)}}  \\ \hline 
  $N_H$&$10^{21}$ см$^{-2}$& 0.4$\pm$ 0.1 \\
 $kT$ & keV& 1.65$\pm 0.01$ \\
 $p$ & & 0.68$\pm0.02$\\
$N_{\rm disk}$ & 10$^{-2}$ & 1.00$\pm0.05$ \\
 $R\cos^{1/2}\theta$ & km & 8.2$\pm 0.2$ \\
 $kT_{\rm e}$ & keV& 12$\pm3$ \\
 $\tau$ & & 1.1$\pm0.2$\\
 $N_{\rm compTT}$ &$10^{-6}$& $9.1\pm0.1$ \\
 $\chi^2_r$, $\chi^2/d.o.f$ & & 0.99, 632.28/638\\ \hline
 \end{tabular} 
\end{table*}

The spectrum is satisfactorily described by the modified accretion disk model \texttt{diskpbb} with
a slope of the power-law temperature distribution in the disk $p\approx 0.65$, a temperature at the inner disk
boundary $kT\approx 2.4$~keV, and an inner disk radius $R\sim 5\cos^{-1/2}\theta$~km. The quality of the fit to the data can be slightly improved by adding a hard component to the modified accretion disk model. A simple power-law component (with a slope $\Gamma\sim 0$), a powerlaw component with an exponential cutoff at energy $\sim 8$~keV, or an accretion disk Comptonization model in a hot corona with a temperature $\approx12$~keV and an optical depth $\tau\sim1$ can be used as the latter with equal success. The measured total X-ray luminosity (0.3--20~keV) of the source is $\sim 2\times 10^{38}$\lum\, from which about $10\%$ is accounted for by the hard 10–--20~keV X-ray band.

\begin{figure*}
\begin{center}
\includegraphics[scale=0.70]{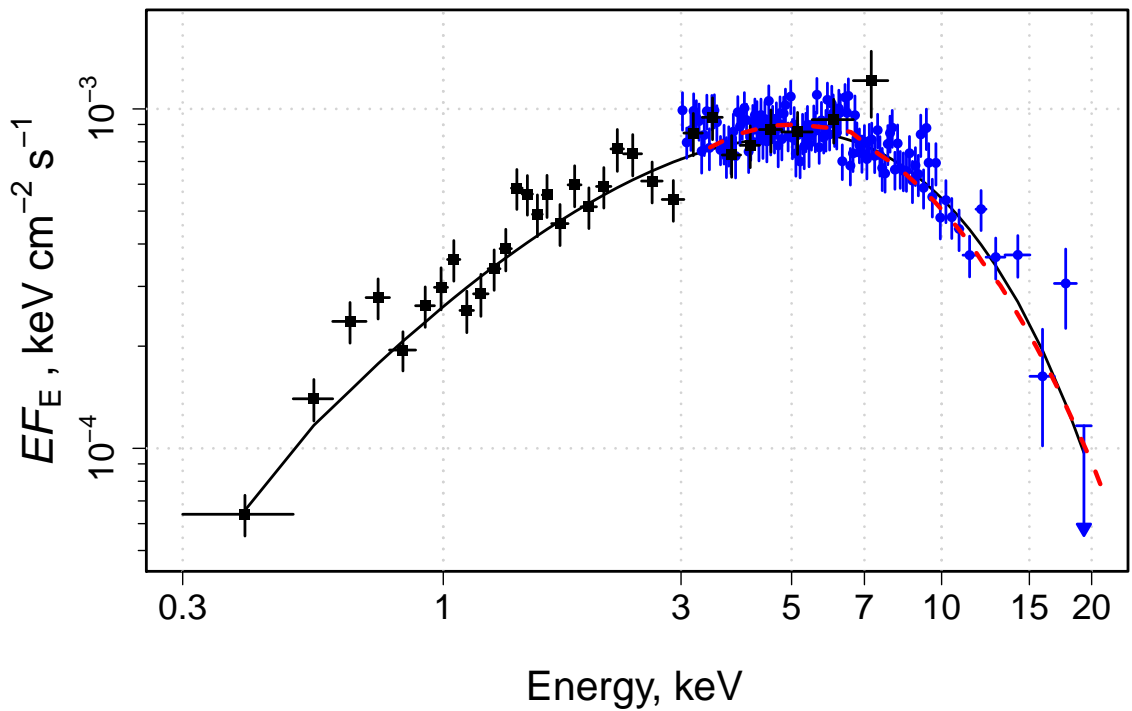}
\caption{ Broadband spectrum of M33 X-6 from the \nu\ and \xrt\ data (similar to Fig.~\ref{ris:all-cutoffpl}). The red dashed line indicates one of the typical spectra for the $Z$-source XTE~J1701--462 \citep{revni2013} with a coefficient of $1.05\times10^{-4}$ for the convenience of comparison.}
\label{ris:z}
\end{center}
\end{figure*}

\begin{figure*}
\begin{center}
\includegraphics[scale=0.55]{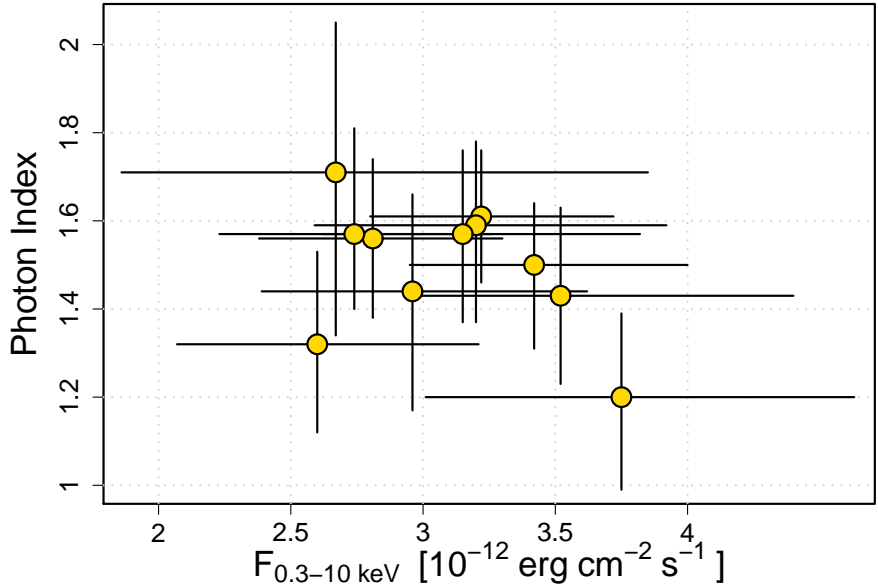}
\caption{Power-law slope (photon index $\Gamma$) versus observed (absorbed) flux from the \xrt\ data in the energy range 0.3--10~keV}
\label{ris:hard}
\end{center}
\end{figure*}

The set of characteristics listed above provides strong evidence that accretion onto a neutron star with a rate approximately equal to the critical one may occur in the binary \x6. Similar X-ray spectra and luminosities have long been observed in our Galaxy for a number of low-mass highluminosity binaries, the so-called $Z$-sources \citep [see][]{hasinger1989}. Indeed, as shown in Fig.~\ref{ris:z}, our spectrum of \x6\ is very similar, for example, to the spectrum of the well-studied $Z$-source XTE~J1701--462 in one of its spectral states \citep{revni2013}. However, it should be noted that the spectra of Z-sources vary noticeably both from source to source and from observation to observation for individual objects (see, e.g., \citealt{lin2009,revni2013}). It is believed that in objects of this type we deal with subcritical accretion onto a weakly magnetized neutron star. The X-ray spectrum of $Z$-sources is usually described by a combination of soft accretion disk emission and hard neutron star boundary layer emission \citet{revni2013,gilfanov2003}, which can be fitted by a power law with a cutoff or Comptonization, as was done in this paper. Thus, the source \x6\ being discussed here can become one of the first candidates for $Z$-sources outside our Galaxy (and its companions). Previously, \cite{barnard2003} detected spectral properties and variability for the source RX~J0042.6+4115 in the galaxy M31 characteristic for $Z$-sources.

However, as has been noted in Subsection~\ref{xrt}, the observed X-ray flux from \x6\ exhibits no strong variability \citep[e.g.,][]{Tullmann11}, which is atypical for X-ray binaries \citep[see, e.g.,][]{remilard2006} and Z-sources, in particular \citep{homan2007}. In searching for possible changes in the spectral hardness of M33 X-6, we fitted the data by a power law with fixed photoabsorption $N_{\rm H} = 1.1 \times10^{21}$~cm$^{-2}$. As follows from Fig.~\ref{ris:hard}, where the photon index is plotted against the observed flux, no pronounced changes in the spectral state were found. However, the errors in the spectral slope and X-ray flux are very large, because the exposures of the \xrt\ observations used are short. We are going to perform a more detailed study of the \x6\ variability in our succeeding paper using additional data from other X-ray telescopes.

\newpage

\section*{ACKNOWLEDGMENTS}
The results presented in this paper were obtained with the Neil Gehrels ({\it Swift}) and \nu\ orbital observatories. This work was supported by the Russian Science Foundation (project no. 14-12-01315).



\end{document}